\documentclass{article}

\PassOptionsToPackage{square,numbers}{natbib}
\usepackage[final]{neurips_2022}

\usepackage[utf8]{inputenc} 
\usepackage[T1]{fontenc}    
\usepackage{hyperref}       
\usepackage{url}            
\usepackage{booktabs}       
\usepackage{amsfonts}       
\usepackage{nicefrac}       
\usepackage{microtype}      
\usepackage{xcolor}         
\usepackage[pdftex]{graphicx}
\usepackage{caption}
\usepackage{subcaption}

\title{Self-driving Multimodal Studies at User Facilities}

\author{%
  Phillip M.~Maffettone\\
  National Synchroton Light Source II\\
  Brookhaven National Laboratory\\
  Upton, NY 11973\\
  \texttt{pmaffetto@bnl.gov}\\
  \And
  Daniel B.~Allan\\
  National Synchroton Light Source II\\
  Brookhaven National Laboratory\\
  Upton, NY 11973\\
  \And
  Stuart I.~Campbell\\
  National Synchroton Light Source II\\
  Brookhaven National Laboratory\\
  Upton, NY 11973\\
  \And
  Matthew R.~Carbone\\
  Computational Science Initiative\\
  Brookhaven National Laboratory\\
  Upton, NY 11973\\
  \And
  Thomas A.~Caswell\\
  National Synchroton Light Source II\\
  Brookhaven National Laboratory\\
  Upton, NY 11973\\
  \And
  Brian L.~DeCost\\
  Material Measurement Lab\\
  National Institute of Standards and Technology\\
  Gaithersburg, MD 20899\\
  \And
  Dmitri Gavrilov\\
  National Synchroton Light Source II\\
  Brookhaven National Laboratory\\
  Upton, NY 11973\\
  \And
  Marcus D.~Hanwell\\
  National Synchroton Light Source II\\
  Brookhaven National Laboratory\\
  Upton, NY 11973\\
  \And
  Howie Joress\\
  Material Measurement Lab\\
  National Institute of Standards and Technology\\
  Gaithersburg, MD 20899\\
  \And
  Joshua Lynch\\
  National Synchroton Light Source II\\
  Brookhaven National Laboratory\\
  Upton, NY 11973\\
  \And
  Bruce Ravel\\
  Material Measurement Lab\\
  National Institute of Standards and Technology\\
  Gaithersburg, MD 20899\\
  \And
  Stuart B.~Wilkins\\
  National Synchroton Light Source II\\
  Brookhaven National Laboratory\\
  Upton, NY 11973\\
  \And
  Jakub Wlodek\\
  National Synchroton Light Source II\\
  Brookhaven National Laboratory\\
  Upton, NY 11973\\
  \And
  Daniel Olds\\
  National Synchroton Light Source II\\
  Brookhaven National Laboratory\\
  Upton, NY 11973\\
  \texttt{dolds@bnl.gov}\\
}

\begin{document}

\maketitle

\begin{abstract}
Multimodal characterization is commonly required for understanding materials.
User facilities possess the infrastructure to perform these measurements, albeit in serial over days to months. 
In this paper, we describe a unified multimodal measurement of a single sample library at distant instruments, driven by a concert of distributed agents that use analysis from each modality to inform the direction of the other in real time.
Powered by the \emph{Bluesky} project at the National Synchrotron Light Source II, this experiment is a world's first for beamline science, and provides a blueprint for future approaches to multimodal and multifidelity experiments at user facilities. 
\end{abstract}

\section{Introduction}

Fully characterizing new materials depends on multiple modalities of measurement, e.g., spectroscopy and diffraction.
Differing modalities often suffer from contrasting fidelity and resource requirements.
A common example of this is the use of X-ray absorption fine structure (XAFS) and total scattering for the characterization of functional materials such as high entropy alloys~\cite{Miracle_2017}, high entropy oxides~\cite{Musico_2020}, and electroactive materials~\cite{Sun_2017}.
Probing these modalities simultaneously and efficiently would accelerate materials analysis, and by extension discovery. 
\par

In many cases, these analysis techniques are only available to researchers at light sources and user facilities. 
Light sources, such as the National Synchrotron Light Source II (NSLS-II) at Brookhaven National Laboratory (BNL), are large scale facilities that provide service science for the research community. 
These are government-funded centers that possess first- or only-in-class measurement capabilities. 
A \emph{beamline} or \emph{end-station} is the instrumentation that provides a measurement capability at a light source. 
Due to advances in both light source accelerator and detector technologies, the productivity of a high-throughput beamline is no longer limited by the amount of photons it can produce and detect, but rather the ability to control and analyze the high rate of measurements.
To help realize and leverage the full potential of these facilities, researchers have automated data collection and integrated artificial intelligence (AI) into the real-time analysis and orchestration of experiments \cite{Campbell_2021, three_pp, Noack_2021}.
In line with this, recent advancements have been made to convert and incorporate beamlines into self-driving labs or materials acceleration platforms (MAPs)~\cite{gamification, Barbour_2022, Seifrid_2022, nsls2_strategy}.

Multimodal measurements are a critical part of NSLS-II capabilities, yet present a further challenge: 
since time is allocated using a proposal system, a sample whose measurement warrants further study using a different modality may have to wait months after analysis to be allocated time on the next instrument.
This bottleneck is exemplified by the \textit{operando} study of a lithium--sulfur battery cell at three different beamlines~\cite{Sun_2017}.
Moreover, the diversity of beamlines and materials studied creates a need for diverse agents to interface with experimental orchestration~\cite{Barbour_2022, nsls2_strategy}. 
Integrating multiple beamlines using distributed agents would not only open the bottleneck for the analysis of engineered devices or combinatorial materials~\cite{Green_2013}, but enable discovery workflows with self-driving synthesis engines. 

In this work, we demonstrate the world's first truly multimodal measurement at a light source. 
The measurement used two physically distant beamlines simultaneously to examine a single sample library in a decentralized control loop. 
Multiple AI agents were able to guide both beamlines in concert, while retaining the opportunity for human experts to engage in decision making.
This work can scale readily to incorporate other beamlines and MAPs, as well as serving as a blueprint for building the experimental orchestration of other MAPs. 

\section{Preliminaries}

\paragraph{X-ray powder diffraction}
X-ray powder diffraction (XRD) is one of the primary characterization methods of the material science and solid state chemistry communities. A diffractometer measures scattered intensity from a sample as a function of angle, from which atomic-scale details can be extracted, such as the material phase, lattice constant, strain, site occupancy, and atomic displacement parameters. At large-scale light source user facilities, measurement times for full XRD patterns are on the order of millisecond to minutes, and are dependent on the sample and instrumentation. At the Pair Distribution Function (PDF) beamline at NSLS-II, it is common to measure a full XRD pattern in 30 seconds~\cite{pdf}.

\paragraph{X-ray Absorption Fine Structure spectroscopy}
X-ray Absorption Fine Structure spectroscopy (XAFS) is another characterization method in wide use in materials science, solid state chemistry, and many other scientific disciplines. In XAFS, an X-ray beam is scanned in energy over a range that includes the binding energy of deep-core electrons in an element contained in the sample. By measuring the change in X-ray absorption cross section as a function of energy, information about the electronic state of the element as well as the local atomic configuration can be obtained.XAFS data were measured at the Beamline for Materials Measurement (BMM), operated by the National Institute of Standards and Technology at NSLS-II. At the BMM beamline at NSLS-II, an XAFS spectrum is typically measured in about 7 minutes~\cite{bmm}.

\paragraph{Measurement task}
The measurement task herein is to fully and efficiently discover and characterize all material phases in a sample library, where the samples vary as a function of coordinate on a single wafer. 
As beamtime at user facilities is a limited resource, optimal collection strategies allow for higher duty cycles, ultimately resulting in more samples being measured and thus, more scientific productivity. 
Therefore, we consider the optimal exploration of the space $\mathcal{D}$, 
which we probe \emph{via} two mappings from two distant beamlines: $f_i: \mathcal{D} \mapsto \mathbb{R}^{n_i}$, 
where  $f_i$ is a single probe of space,
$\mathcal{D}$ is the continuous space of positions in a wafer library $\Sigma$,
and $n_i$ is the data length of a given spectral measurement. In each phase of measurement we select $k$ new points  $D_{\mathrm{query}} \subset \mathcal{D}$ at which to query $f_i$. The goal is to find all unique values of $f_i$ with the fewest number of queries.

\section{Decentralized control of beamlines}
We rely on two broad advancements to enable multimodal analysis MAPs. 
Firstly, we deployed industry-standard technology to limit isolation, improve system reliability, and scale. 
Secondly, we focused on moving beyond closed loop control for experiment orchestration. 
These advancements apply to almost all 28 operational beamlines at NSLS-II, and will apply to the 22 more expected to be built. 

\paragraph{Infrastructure developments for scalability}
In a recent multi-million dollar effort, BNL has built a \textit{High Throughput Science Network} that connects NSLS-II to BNL super computers and research centers~\cite{nsls2_strategy}. 
This included a reprovisioning of the IP space of NSLS-II that enabled firewalled communication between beamlines, the data storage center, and compute for data processing. 
Virtualization was used to deploy all services in this work using VMWare\footnote{\label{foot:disclaimer}Any mention of a commercial product is for information purposes only. It does not imply any recommendation or endorsement by the National Institute of Standards and Technology (NIST) or Brookhaven National Laboratory (BNL).} clusters run centrally and at the beamlines~\cite{vmware}. This enabled increased fault tolerance and rapid prototyping of tools.
Redundant MongoDB services~\cite{mongo}  were deployed for document-style data storage and retrieval, with a central Lustre~\cite{lustre} file store for most large data. 
Data access was largely achieved using Tiled~\cite{tiled}, which is a web-based storage agnostic tool replacing DataBroker~\cite{bluesky}.  
A Kafka service~\cite{kafka} was deployed that published the document stream of all beamline measurements, which were then subscribed to by agents or other services. 
At a facility scale the Ansible  automation platform is used to provision VMs and services~\cite{ansible}. 
Lastly, secure remote access to beamlines was acheived using Guacamole~\cite{guacamole}, with data access and remote analysis provided by a JupyterHub~\cite{jupyter_hub}.

\paragraph{Experimental orchestration using the \emph{Bluesky} suite}
The \emph{Bluesky} project is a collection of Python libraries for experimental science with thousands of users and nearly 100 community developers working on active forks. While co-developed, each package is designed to be independently used~\cite{bluesky}.
A beamline consists of many devices (motors, detectors, pumps, sensors, etc.) that can be orchestrated to conduct experimental plans as asynchronous coroutines using the \emph{Bluesky Run Engine}. 
Current `autonomous experiments' at user facilities operate in synchronous closed loop measurements over a single modality~\cite{Noack_2021}.
This lock-step approach to experiment and analysis, leaves no room for human experts to engage in the loop, or the incorporation of information from  complementary techniques. 
\par

\begin{figure}
  \centering
  \begin{subfigure}{0.49\linewidth}
     \centering
     \includegraphics[width=\textwidth]{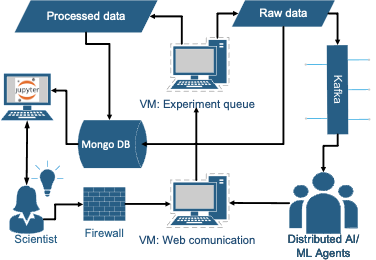}
 \end{subfigure}
 \hfill
 \begin{subfigure}{0.49\linewidth}
     \centering
     \includegraphics[width=\textwidth]{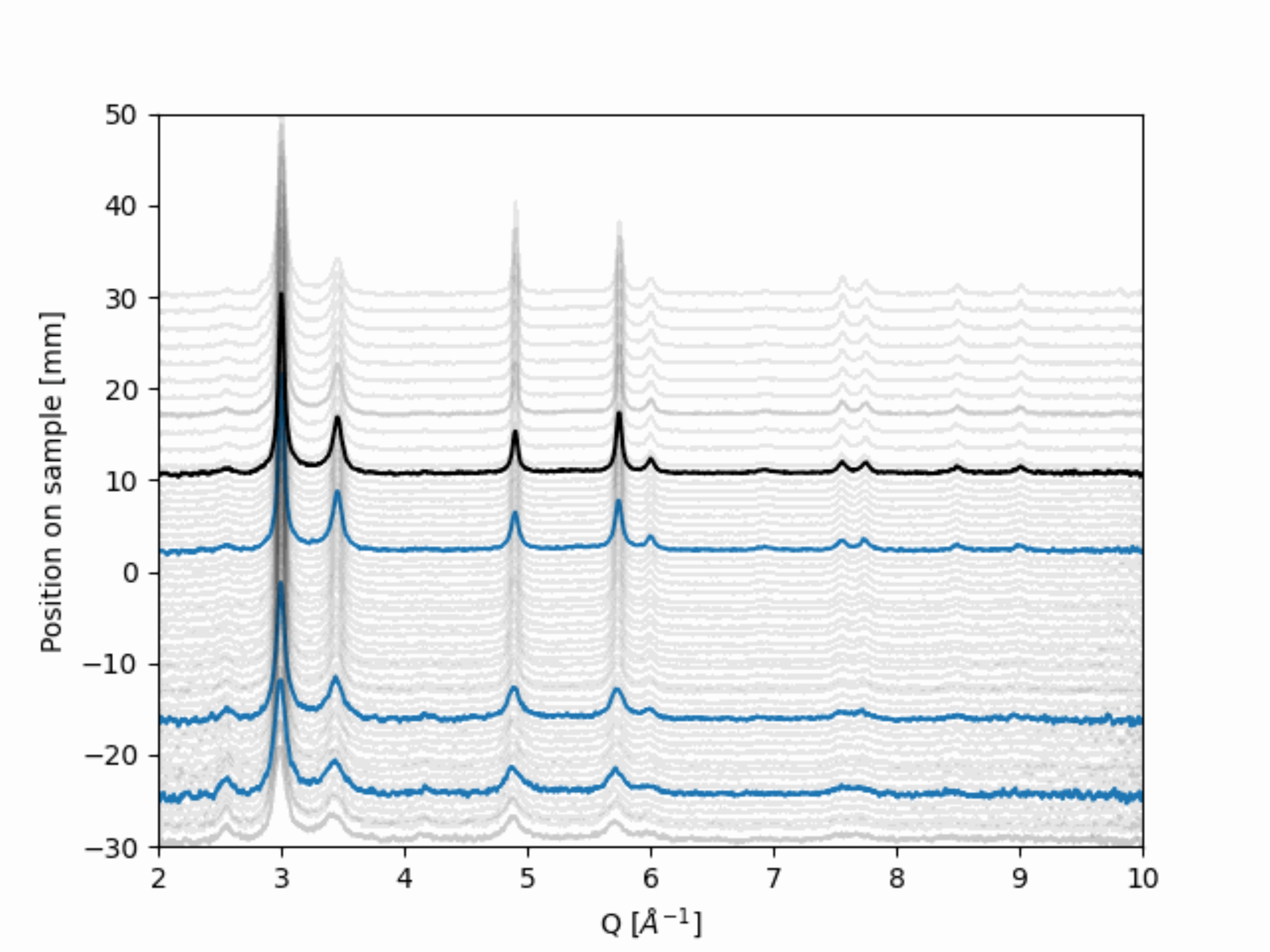}
 \end{subfigure}
 \hfill
  \caption{Left: block diagram for decentralized orchestration of a single beamline. Each beamline can be linked \emph{via} agents with access to the respective VM for web communication (http-server), Kafka node, and/or database. 
  Right: a snapshot from the commissioning experiment. The PDF monarch read all historical XRD data (grey), suggested the next point in a geometric design of experiments for XRD (black), and periodically suggested regions of maximal explained variance for further study using XAFS (blue).}
  \label{fig:scheme}
\end{figure}

With recently developed packages in the \emph{Bluesky} project and the infrastructure advancements at NSLS-II, we first executed decentralized control of a single beamline using artificial intelligence (AI) and human agents.
A queue server~\cite{queue_server} replaced the command line execution of plans by the \emph{Run Engine} with a mutable priority queue. 
The queue server manages permissions, and was securely accessed by human and AI agents through http protocols using the http-server~\cite{http_server}, and monitored through a graphical user interface. 
A dedicated queue server was run on a VM within a beamline subnet, with dedicated http-servers run on VMs on a central network. 
As the beamline executed plans, it published raw data to a Kafka topic and wrote to a central database. 
Services subscribed to the raw data stream and produced processed data that was concurrently stored and referenced in the database. 
Agents then accessed the raw or processed data to produce visualizations or suggest follow up experimental plans using the http-server.
There was no limit to the number of agents, or their location. 
In this work, AI agents were deployed on physical machines within the NSLS-II IP space, and human agents engaged remotely.
In Figure~\ref{fig:scheme}, we show the block diagram of a single self-driving beamline alongside a snapshot of the commissioning experiment as seen by one agent. 

\section{Agent driven multimodal characterization}
In this work, we combined the decentralized experimental orchestration of two distant beamlines \emph{in concert}. Together, both beamlines were exploring an identical sample library, seeking to extract the most information from the library. 
The control system included passive agents for data processing, unsupervised agents that combined a design of experiments approach and entropy search, pre-trained deep ensembles, and human experts. 
This simultaneous combination of beamlines demonstrates a world first for synchrotron science, and lays the framework for future experiments using more specialized agents and complex experimental spaces. 

\paragraph{Agent design}
The agents designed for this experiment considered what was rational, time sensitive, and interesting. 
The agents were developed using an \texttt{ask}--\texttt{tell}--\texttt{report} grammar, where an agent could be `told' about new data, `asked' what to do next, and `report' about it's current status~\cite{three_pp}. 
Each step was recorded using the streaming event model of \emph{Bluesky}, \emph{via} \emph{Tiled}, so that the agent's perspective and decision making could be played back for investigation. 
\par

\begin{figure}[h]
  \centering
  \includegraphics[width=\linewidth]{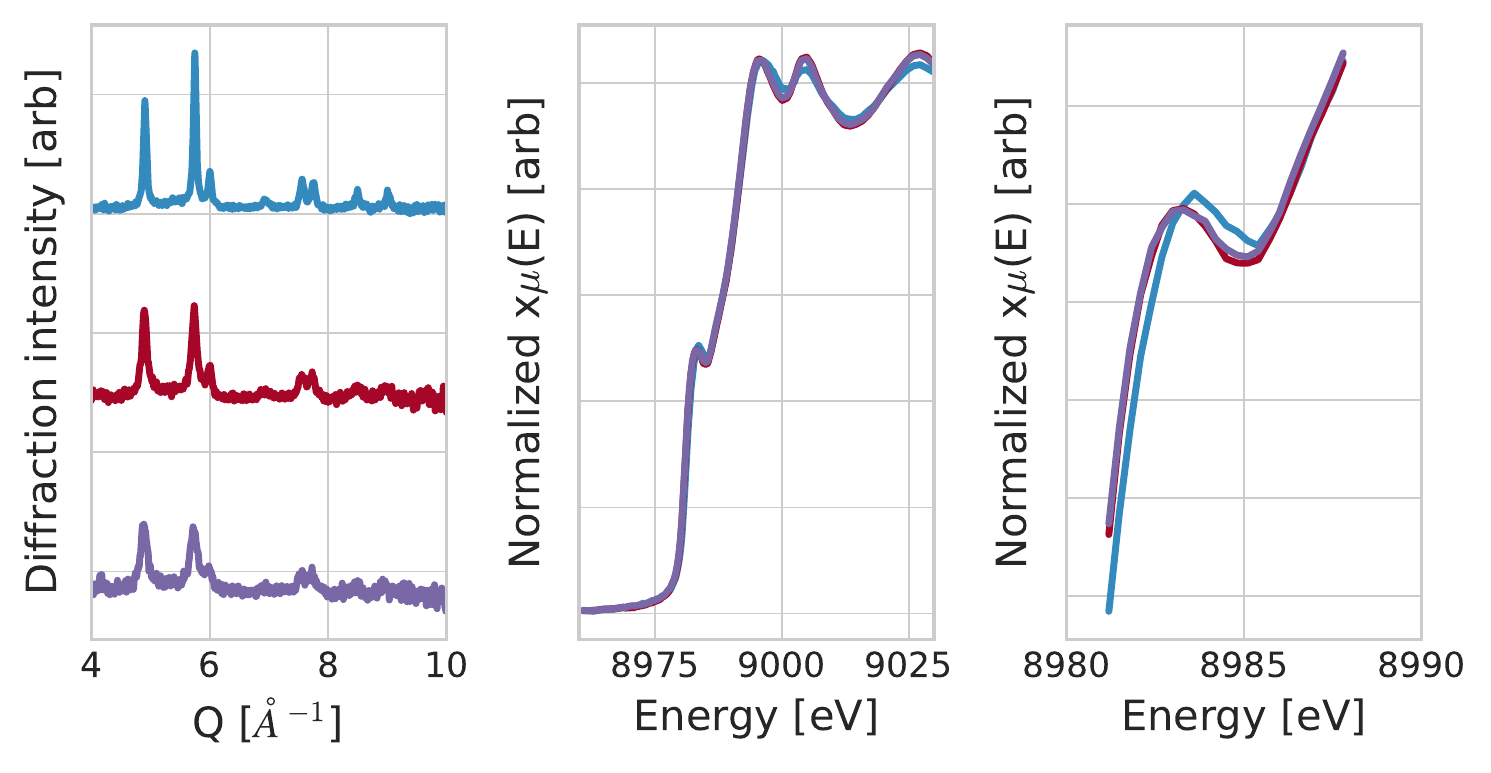}
  \caption{Measurements suggested by the PDF Monarch. The changes between the regions with increasing Cu content were extremely subtle, which drove the consideration of unsupervised techniques over pretrained deep models that were conditioned on more phases. Left: The XRD of the three phases over a Q-range that shows short range order changes in the patterns. 
  Middle: The XAFS spectra showing the the X-ray absorption near edge structure (XANES). 
  Right: The leading edge of the spectra showing the changing shoulder. }
  \label{fig:spectra}
\end{figure}

We used a monarch--subject relationship to let an agent subscribing to one beamline dictate the plans of the opposite beamline.
At baseline, an agent would propose a design of experiments approach for a  measurement on the queue. 
The monarch–subject agents gave the “monarch” agent priority to dictate the next measurements on the “subject” queue at some interval. 
For example, the PDF-monarch consumed all the XRD data measured at PDF, and every hour would dictate some regions of interest for further analysis at the BMM-subject. 
Because the PDF measurement was much faster than BMM, the BMM-monarch would dictate at every round of design. 
Regions of interest were determined \emph{via} maximal explained variance using alternating K-means clustering and constrained matrix factorization~\cite{cmf}.
A snapshot of the PDF monarch is shown in Figure~\ref{fig:scheme} with the subject spectra resulting from its decisions detailed in Figure~\ref{fig:spectra}.
Passive agents were processing and visualizing  data from the experiments for expert interpretation.

\par 

Additional AI agents were built for the experiment, albeit were unused.
These included deep feed-forward ensemble models that consumed 1-d spectra and predicted the corresponding crystallographic phase, and convolutional variational autoencoders to compress the spectral space into a lower dimensional subspace~\cite{xca, Banko_2021}. 
These models were pre-trained on synthetic data produced using the computational equilibrium phase diagram from the Materials Project \cite{materials_project} according to the methods of \citet{xca}.
Still using the monarch--subject paradigm, the ensemble model was designed to suggest regions of maximum information entropy.

The full space of potential features that could be measured in the 1D spectra is not necessarily known \textit{a priori}, which limits the reliability of pre-trained classification tools. We used a variational autoencoder (VAE) to compress the raw data into a latent representation of the information, with some tolerance to address unexpected or novel features \cite{Banko_2021}.
As developed, the experiment can be autonomously guided using a Gaussian process that was continually retrained to predict the latent representation of the VAE, and then sampled with Bayesian optimization using an entropy search.
Unfortunately, during the commissioning study we found the data from the physical sample library was out of distribution of the synthetic data used to train the deep feed-forward models and VAE.
The distribution of the training data is defined by the phase data from the Materials Project and the physics informed data augmentation parameters used in the simulation. 
Thus, only the naive and unsupervised models were employed for control of the experiment.

\paragraph{Human in the loop}
Scientists were able to interact with the self-driving control loop through raw and processed data access, inspection of agent analysis, and direct manipulation of the experimental queue (Fig.~\ref{fig:scheme}).
Throughout the experiment, experts accessed the measurement data in real time using \emph{Tiled}, and suggested new measurements based on their manual analysis.
Since each \texttt{ask}, \texttt{tell}, and \texttt{report} by an agent was published to a MongoDB, the scientists were able to probe the AI-driven decisions and reports with the same interface, and used that information to drive human decision making. 
Agents were modified by restarting processes with new hyperparameters and retrained using the experimental history, or by directives using a Kafka interface.
A human could add suggestions to the queue server using a graphical user interface (Queue Monitor) or the Python API. 
Combining a federation of agents with human engagement in the loop  offers better insights and efficiencies, \emph{e.g.,} by visualizing the deep network's predictions we interpreted the out of distribution failure mode and prohibited the agent from adding suggestions to the queue. 

\paragraph{Commissioning example}
For this initial commissioning study, we selected a sample library that was both readily characterized by the two techniques and assuredly identical in composition. A bimetallic alloy (CuTi) was cast onto a circular Si wafer, where the variable composition was dependent on the position on the wafer. This resulted in a wafer with  a high concentration of Cu on one side, a high concentration of Ti on the other, and a presumed linearly varying combination of the two across the wafer. 
Although the fabrication is reproducible, we ensured identical samples by cleaving the prepared circular wafer in half, parallel to the direction of the deposition gradient. Thus, the composition of the sample could be considered functionally identical at either half of the cleave.
The wafers were mounted on standard stages on both beamlines, such that they could be independently positioned with the instrument stage to any point within 1 to 5 seconds. 
\par 


\section{Discussion and Future Work}
While our initial commissioning experiment demonstrates a unique capability for beamline science, there are several avenues for improvement.
From an infrastructure standpoint, services that were deployed using virtualization (e.g., the queue-server), could be reproducibly developed, version controlled, and efficiently deployed using containerization.
Furthermore, integrating synthesis MAPs that are not natively developed using \emph{Bluesky}, would unlock on-the-fly synthesis and characterization of dynamic sample libraries. 
To this end, more complex experimental spaces will be explored in the future, which will in turn warrant the integration of more complex agents. 
\par

Since the entropy search developed in this initial experiment struggled with an inadequate compression of the true experimental space by the synthetically trained VAE, it would be beneficial to have more flexible measures of scientific information to search over in the future.
Furthermore, methodological work is needed to build physics-aware agents that can effectively combine both the XRD and XAFS data, and use information from both measurements to drive both measurements.
Lastly, we propose the need for meta-agents, or adjudicators, which coordinate between a collection of agents in more sophisticated ways than a simple priority queue and provide an additional avenue for human intervention.
\par

In summary, we demonstrated the capability and blueprint to conduct self-driving multimodal, multifidelity analysis using multiple beamlines. 
Using decentralized control of two self-driving beamlines, we allowed for multiple agents---both computational and human---to contribute to the experimental planning. 
We commissioned this technology in the study of a well understood bimetallic alloy library, with a more complex library studied in November 2022.
This work opened up new opportunities for high-throughput user science that studies more complex multimodal and multifidelity tasks, and the development of physics-imbued multimodal AI agents.
We expect facile incorporation of new beamlines at NSLS-II or other facilities using \emph{Bluesky} (e.g., the combined neutron and X-ray measurement of a system).
This world's-first marks a step forward in making optimal use of light sources and user facilities.

\begin{ack}
This research used the PDF, and BMM beamlines of the National Synchrotron Light Source II, a U.S. Department of Energy (DOE) Office of Science User Facility operated for the DOE Office of Science by Brookhaven National Laboratory (BNL) under Contract No. DE-SC0012704, and resources of a BNL Laboratory Directed Research and Development (LDRD) projects 
23-039 "Extensible robotic beamline scientist for self-driving total scattering studies", 20-032 “Accelerating materials discovery with total scattering via machine learning” and 22-059 "Precision synthesis of multiscale nanomaterials through AI-guided robotics for advanced catalysts."
\end{ack}

\section*{Author contribution}
D.O. proposed the project.
P.M.M., B.R., and D.O. conceived and developed the commissioning experiment.
B.L.D. and H.J. suggested and created the material sample studied.
P.M.M. and M.R.C. developed the agents. 
P.M.M. and T.C.C. deployed agents and Bluesky adaptive tooling. 
M.D.H. and S.C.C. deployed and maintained critical infrastructure. 
J.L. managed the Kafka deployment.
D.A. led the development of Tiled for data access.
D.G. led the development of Bluesky Queueserver and related technology for distributed experiments.
J.W. supported beamline hardware and EPICS integration.
S.W. oversaw network and security infrastructure. 
B.R. managed the operation and integration of the BMM beamline.
D.O. managed the operation and integration of the PDF beamline. 
P.M.M. wrote the manuscript with input from all authors.

\bibliography{bibliography.bib}

\appendix

\section{Code availability}
All of the code for driving the experiments and developing the models is available at \href{https://github.com/NSLS-II-PDF/mmm-experiments}{github.com/NSLS-II-PDF/mmm-experiments} under the BSD 3-Clause license. This is a repository under active development, with the last commit in line with the described work at \href{https://github.com/NSLS-II-PDF/mmm-experiments/tree/2299acc11bb911e72e68e4b658d022cc775b5868}{2299acc11bb911e72e68e4b658d022cc775b5868}.

Similarly, the underlying components are developed under the BSD 3-Clause license as parts of the \emph{Bluesky} project. For security reasons, database and network configurations are not publically available.

\section{Agent training}
Code and details for training the deep ensemble models and VAE are available in the above repository. All models were trained continuously using live data synthesis for ~1 week prior to the experiment on an NVIDIA A1000 GPU. 
The crystallography companion agent package was used, XRD parameters tabulated below.
\begin{table}[h]
    \centering
        \caption{Parameters used for generating synthetic XRD patterns from crystallographic structures. Single values were fixed in the calculation, whereas ranges were uniformly sampled in generating each new pattern.}
    \begin{tabular}{c|c}
        Parameter & Range \\ \hline
         wavelength\,[\AA]&0.1655  \\
         noise std&5e-4\\
         instrument radius\,[mm]& 1000.0\\
         theta\_m\,[deg]& 0.0\\
         min 2$\theta$\,[deg]&0.1\\
         max 2$\theta$\,[deg]&12.0\\
         num datapoints&3000\\
         background 0th order & (0, 1e-3)\\
         background 1st order & (-1e-4, 1e-4)\\
         background 2nd order & (-1e-4, 1e-4)\\
         march parameter & (0.5, 1.0) \\ 
         fractional isotropic expansion  & (-0.05, 0.05)\\

    \end{tabular}
    \label{tab:my_label}
\end{table}
\par
The feed forward ensemble architecture consisted of 25 independent convolutional neural networks (CNN), each with 3 layers of convolution. The convolutional layers had [8,8,4] channels respectively, each with a kernel size of 5 and stride of 2. 
The VAE consisted of a convolutional encoder-decoder network with a 2 dimensional latent space for ease of visualization. The decoder was automatically generated to match the encoder impact on data shape. The encoder consisted of 2 CNN layers with filters of [8, 4], kernel sizes of [5, 5], strides of [2, 1], and max pooling of [2, 2] respectively, followed by a dense layer to compress the output to the latent dimensions.
The agents were deployed on a single workstation with a 32 core CPU during the experiment, though not engaged for suggestions on the experimental queue.

\end{document}